\documentclass{ws-rv961x669}
\usepackage{ws-rv-van}     
\usepackage{ws-rv-thm}     
\usepackage{subfigure}     
\usepackage{graphicx}
\usepackage{mathrsfs}
\makeindex
\def\beq{\begin{equation}}
\def\eeq{\end{equation}}

\newcommand{\bo}{\raise-1mm\hbox{\Large$\Box$}}

\newcommand{\f}[2]{\frac{#1}{#2}}

\newcommand{\la}{\langle}
\newcommand{\ra}{\rangle}
\newcommand{\w}{\omega}
\newcommand{\kp}{\kappa}
\newcommand{\be}{\begin{equation}}
\newcommand{\ee}{\end{equation}}
\newcommand{\bea}{\begin{eqnarray}}
\newcommand{\eea}{\end{eqnarray}}

\renewcommand{\d}[1]{\ensuremath{\operatorname{d}\!{#1}}}
\begin{document}

\chapter{Reflecting at the Speed of Light}
\author{Michael R.R. Good}
\address{Department of Physics, School of Science and Technology,\\
Nazarbayev University, Astana, Republic of Kazakhstan\\
E-mail: michael.good@nu.edu.kz}


\author{\textit{In memory of Kerson Huang}}

\begin{abstract}
A perfectly reflecting accelerating boundary produces thermal emission to an observer at $\mathscr{I}_L^+$ and a finite amount of energy to an observer at $\mathscr{I}_R^+$ by asymptotically traveling to the speed of light without an acceleration horizon.  
\end{abstract}


\body


\vspace{0.5cm}
Kerson Huang's most recent interest in the connections between quantum theory and gravitation led to an investigation of relativistic superfluidity \cite{Xiong:2014oga} and geometric creation of quantum vortices \cite{Good:2014iua}.  The productive enterprise of quantum field theory under external conditions such as these and others (curved spacetime, moving mirrors, expanding cosmologies, accelerating world lines, etc.) has continued to give interesting results and provoke puzzling questions\footnote{e.g. are black holes springlike?\cite{Good:2014uja} Rotating black holes have temperature $2\pi T = g - k$, where $g = (4M)^{-1}$ is the Schwarzschild surface gravity and $k=M\Omega^2$ is the spring constant.}.  The following note outlines one such exactly solvable external condition, -- a moving mirror; which travels to the speed of light, producing both Planckian distributed particles to $\mathscr{I}_L^+$, and finite energy emission to $\mathscr{I}_R^+$.  It is dedicated to the memory of Kerson and his inspiring work and enlivening curiosity. \\ 

\begin{figure}[ht]
\begin{minipage}[b]{0.45\linewidth}
\centering
 \rotatebox{90}{\includegraphics[width=\textwidth]{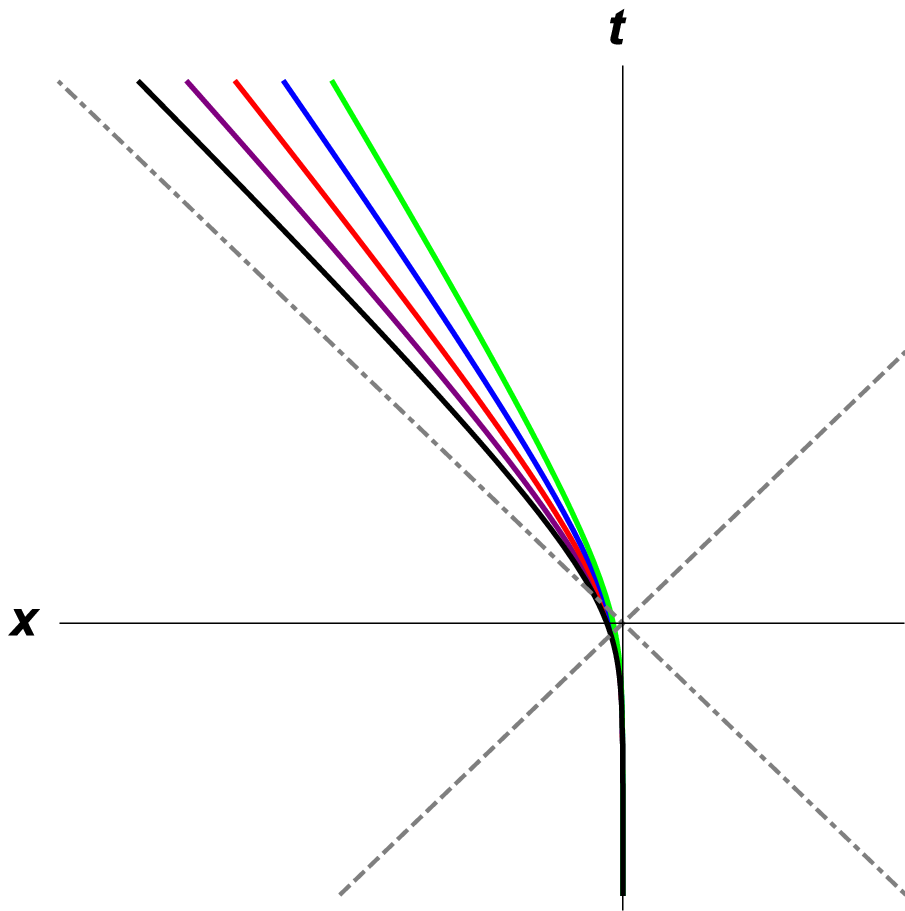}}
\caption{\label{Fig1} Here we graph the mirror trajectory, Eq.~(\ref{dproex}), with various drifting speeds, $s=0.6,0.7,0.8,0.9,1.0$.  $\kp=1$.}
\end{minipage}
\hspace{0.5cm}
\begin{minipage}[b]{0.45\linewidth}
\centering
\includegraphics[width=\textwidth]{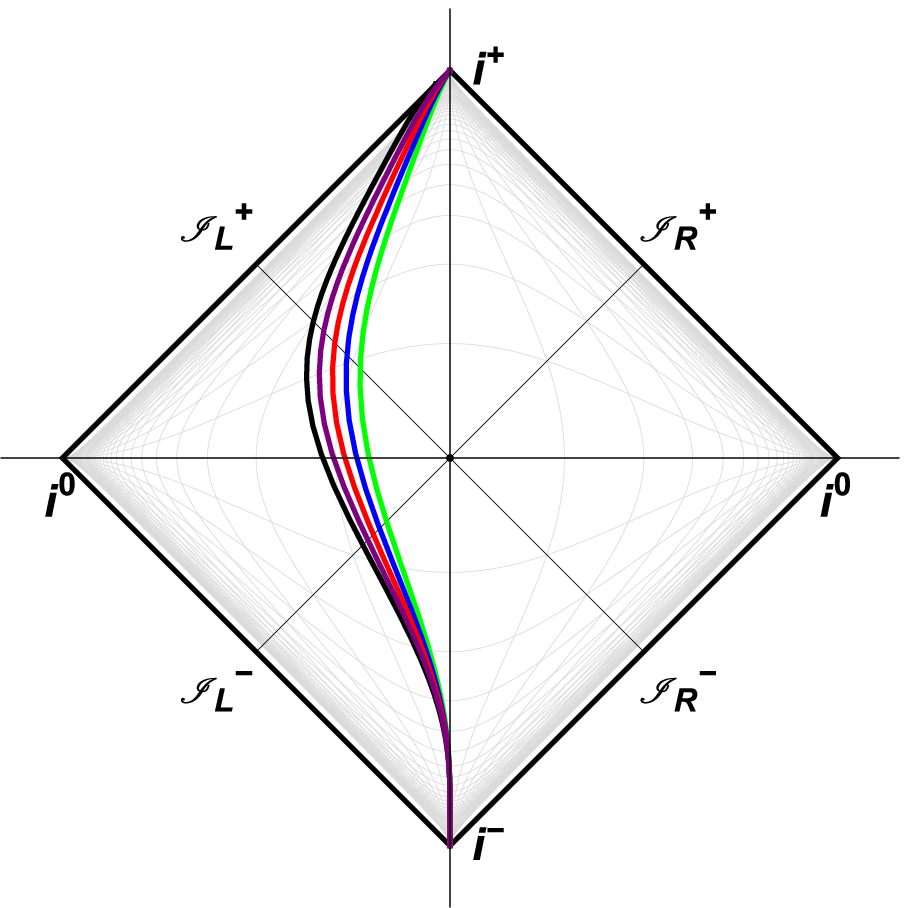}
\caption{\label{Fig2} Eq.~(\ref{dproex}), in a Penrose diagram with drifting speeds $s=0.6,0.7,0.8,0.9,1.0$.  Even when $s=1$, the mirror has no horizon. $\kp = 1$. }
\end{minipage}
\end{figure}

\begin{figure}[ht]
\begin{minipage}[b]{0.45\linewidth}
\centering
\includegraphics[width=\textwidth]{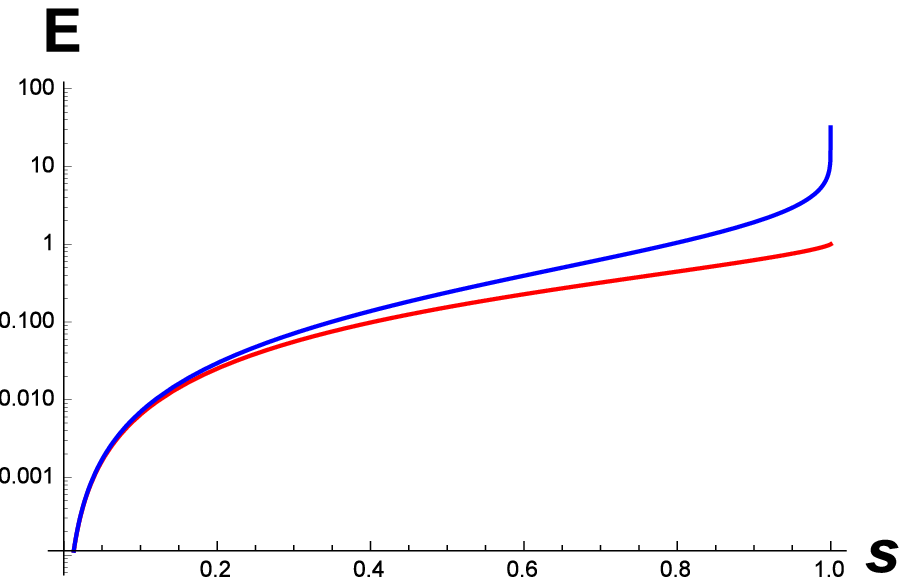}
\caption{\label{Fig3} Log-linear plot of the energy vs. drifting speed of the divergent global energy to the left observer and finite global energy to the right observer. $\kp = 96\pi$, so that $E_R=1$ for $s\rightarrow 1$.   }
\end{minipage}
\hspace{0.5cm}
\begin{minipage}[b]{0.45\linewidth}
\centering
\includegraphics[width=\textwidth]{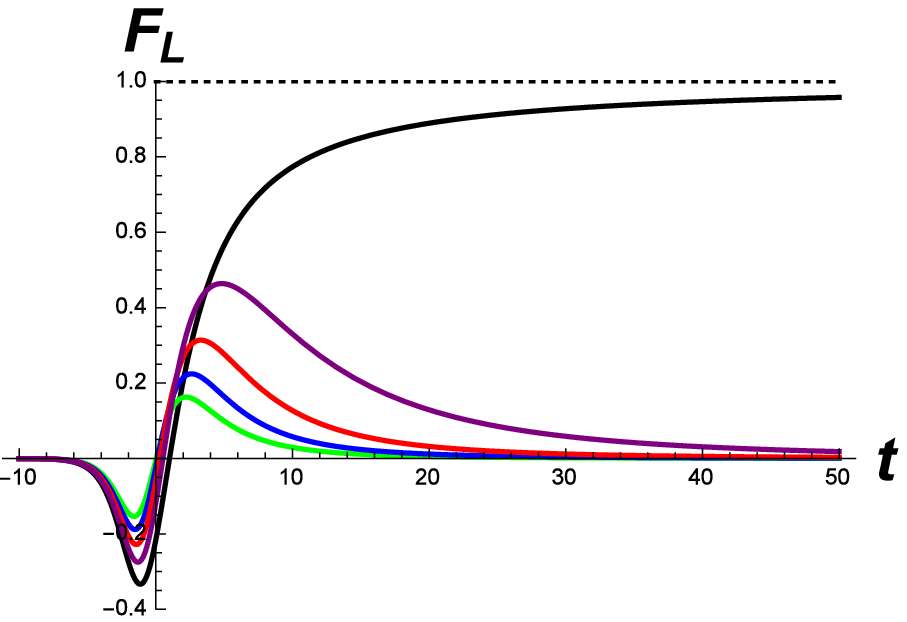}
\caption{\label{Fig4} The approach of energy flux to constant equilibrium emission to the left observer.  $F_{\rm{max}} =1$ for $s\rightarrow1$. The various drifting speeds are the same as the spacetime diagram. $\kp^2 = 48\pi$. }
\end{minipage}
\end{figure}

Mirrors that accelerate forever, asymptotically reaching the speed of light, can produce thermal emission\cite{Davies:1976hi} \cite{Davies:1977yv},
\be |\beta_{\w \w'}|^2 = \frac{1}{2\pi  \kappa \omega' }\frac{1}{ e^{2\pi \kappa/\omega}-1}. \label{thermalB2}\ee
but those that do also produce infinite energy.\footnote{Despite infinite energy, the spin-statistics theorem survives the acceleration horizon forming dynamics \cite{Good:2012cp}.}  Is it possible to \textit{decelerate} a mirror all the way to the speed of light, yet still achieve thermal emission?  Does this render the energy finite? The answer is yes.  
It is interesting to consider the mirror trajectory given by the motion:
\be x(t) = - \frac{s}{\kappa} W(e^{\kappa t}), \label{dproex} \ee
plotted in Figures (\ref{Fig1}) and (\ref{Fig2}), and the inverse, 
\be t(x) =\frac{1}{\kp} \log \left(-\frac{\kappa}{s}  x e^{-\frac{\kappa  x}{s }}\right). \label{timespace} \ee
The $W$ is the product log, which also shows up in the eternally thermal mirror \cite{Carlitz:1986nh, Good:2013lca} and the black mirror\cite{Good:2016oey, Good:2016bsq, paper1, paper2}.  The acceleration parameter is $\kp>0$.  The final coasting speed, $\lim_{t\rightarrow \infty} |\dot{x}(t)| = s$, is between $0<s\leq 1$. 
The motion starts asymptotically static and ends asymptotically drifting.  To compute the total energy emitted to the right observer at $\mathscr{I}_R^+$, one can first find the energy flux \cite{Good:2016atu}:  
\be F(x) = \frac{1}{12\pi}\left[\f{t'''(x)(t'(x)^2-1)-3t'(x)t''(x)^2}{(t'(x)-1)^4(t'(x)+1)^2}\right], \ee
which gives, for trajectory Eq.~(\ref{timespace}) to $\mathscr{I}_R^+$,
\be F_R(x) = -\frac{\kappa ^3 s ^4 x \left(s ^2+2 \kappa ^2 \left(s ^2-1\right) x^2+\kappa  s  x\right)}{12 \pi  (s +\kappa  (s -1) x)^2 (s -\kappa  (s +1) x)^4}. \label{FR}\ee
Then, the energy can be calculated from \cite{Good:2016atu}
\be E_R = \int_{0}^{-\infty} F_R(x) (t'(x) - 1)\d x. \ee 
The bounds are such because the mirror starts at $x=0$ and moves left, eventually to $x\rightarrow -\infty$.  The result is,   
\be E_R(s) = \left(\frac{3-2s}{s} +(s-1)\frac{(3+s)\eta}{s^2}\right)\frac{\kappa}{96\pi}, \label{energyright} \ee
where $\eta$ is the final drifting rapidity, $\eta = \tanh^{-1}s$.  This is a monotonic form similar to a pulsed energy flux trajectory\cite{Good:2015nja} but there the energy diverges when $s \rightarrow 1$. Here the energy is finite as $s\rightarrow 1$:
\be E_R = \frac{\kappa}{96\pi}. \ee
To find the total energy to the left observer at $\mathscr{I}_L^+$, one can switch signs on the final drifting speed in Eq.~(\ref{energyright}), $s \leftrightarrow -s$, by symmetry.\footnote{As the energy $E_L$ is concerned, as long as the drift is $s<1$, then $\eta$ is finite, and $E_T$ remains finite.  }  The total energy emitted to both sides is $E_T = E_R + E_L$, 
\be E_T = \frac{\kappa}{24\pi}\left(\frac{\eta}{s} - 1\right). \ee
Note that $E_T>0$ because the drifting rapidity is always larger than the drifting speed, $\eta > s$ for $0< s \leq 1$.  Both $E_R(s)$ and $E_L(s)$ are plotted in Fig.~(\ref{Fig3}). 

To calculate the particle spectrum observed by the right observer at $\mathscr{I}^+_R$, one can find the beta Bogolubov coefficients from \cite{Good:2016atu}
\be \beta_{\w\w'} =
\f{1}{4\pi\sqrt{\w\w'}}\int_{0}^{-\infty} \d x \;
e^{i\w_{n} x -i\w_{p}t(x)} \left(\w_{p} - \w_{n}t'(x) \right) \label{betat}\;.
\ee
where $\w_{p} \equiv \w + \w'$ and $\w_{n} \equiv \w -\w'$. The result is
\be \beta_{\w \w'}(s) = \frac{-i}{\pi} s^{i \omega_p} \sqrt{\omega\omega'} \left(i \w_s \right)^{i\omega_p -1} \Gamma(-i \omega_p),\ee
where $\w_s \equiv (s^{-1} +1)\omega + (s^{-1}-1)\omega'$.  The spectral mode density is
\be |\beta_{\w \w'}(s)|^2 = \frac{\omega'}{\pi  \kappa  \left(\omega' + \omega \right)}\frac{2 \omega }{\w_s^2}\frac{1}{ e^{\frac{2\pi}{\kappa}  \left(\omega + \omega'\right)}-1}.\ee
In the limit the mirror coasts to the speed of light, $s \rightarrow 1$, this is
\be |\beta_{\w \w'}|^2 = \frac{\omega'}{2\pi  \kappa \omega  \left(\omega' + \omega \right)}\frac{1}{ e^{\frac{2\pi}{\kappa}  \left(\omega + \omega'\right)}-1}. \label{proexB2}\ee
The particle spectrum,
\be \la N_\omega \ra = \int_0^\infty |\beta_{\omega\omega'}|^2 \d \w , \ee
at this speed using Eq.~(\ref{proexB2}) is analytically tractable.  It is, 
\be \la N_\omega \ra = -\frac{1}{4\pi^2 \omega} \ln (1- e^{-2 \pi \omega/\kappa}) + \frac{1}{2\pi \kappa} \sum_{m=1}^{\infty} E_1(-\frac{2\pi \omega m}{\kappa}), \ee
where $E_1$ is the exponential integral.  In the high frequency limit, where $\w'\gg \w$ using Eq.~(\ref{proexB2}), the mirror emits particles to the right observer at $\mathscr{I}^+_R$ with spectra 
\be |\beta_{\w \w'}|^2 = \frac{1}{2\pi  \kappa \omega }\frac{1}{ e^{2\pi \kappa/\omega'}-1}. \label{thermalleft}\ee
To the observer at $\mathscr{I}^+_L$, this is thermal emission, $T=\kappa/2\pi$, Eq.~(\ref{thermalB2}), with $\w\leftrightarrow \w'$. Using Eq.~(\ref{FR}), it can be seen that $\mathscr{I}^+_L$ detects constant energy flux emission $F_L = \kappa^2/48\pi$ as $s \rightarrow 1$ by reversing the signs $s\leftrightarrow -s$.  This flux is plotted in Fig.~(\ref{Fig4}). 


%
%
%
%
%
%
%
%
%
%
%
%
%

%

\section*{Acknowledgments}

MG thanks Xiong Chi, K.K. Phua, Yen Chin Ong, Paul Anderson and Charles Evans.

\end{document}